\documentstyle[prb,aps,twocolumn]{revtex}
\def\de{\dot{e}}
\def\du{\dot{u}}
\def\cf{{\cal F}}
\def\cg{{\cal G}}
\def\bfr{{\bf r}}
\def\bfu{{\bf u}}

\input epsf
\epsfverbosetrue
\tighten
\begin{document}
\draft
\twocolumn[\hsize\textwidth\columnwidth\hsize\csname @twocolumnfalse\endcsname

\title{Simulations of cubic-tetragonal ferroelastics}

\author{A. E. Jacobs$^1$\cite{Allan}, S. H. Curnoe$^2$\cite{Stephanie} 
and R. C. Desai$^1$\cite{Rashmi}}
\address{$^1$Department of Physics, University of Toronto,
Toronto, Ont., M5S 1A7, Canada\\
$^2$Department of Physics and Physical Oceanography, Memorial University of
Newfoundland,\\ St. John's, Nfld.\ \& Lbdr., A1B 3X7, Canada}

\date{\today}

\maketitle

\begin{abstract}
We study domain patterns in cubic-tetragonal ferroelastics by solving 
numerically equations of motion derived from a Landau model of the phase 
transition, including dissipative stresses.
Our system sizes, of up to $256^3$ points, are large enough to reveal many 
structures observed experimentally. 
Most patterns found at late stages in the relaxation are multiply banded; 
all three tetragonal variants appear, but inequivalently. 
Two of the variants form broad primary bands; 
the third intrudes into the others to form narrow secondary bands with 
the hosts. 
On colliding with walls between the primary variants, the third either 
terminates or forms a chevron. 
The multipy banded patterns, with the two domain sizes, the chevrons and the 
terminations, are seen in the microscopy of zirconia and other 
cubic-tetragonal ferroelastics. 
We examine also transient structures obtained much earlier in the relaxation; 
these show the above features and others also observed in experiment. 

\end{abstract}

\pacs{PACS numbers: 81.30.Kf, 62.20.Dc}
]
\narrowtext
\tightenlines

\section{Introduction}
Ferroelastics\cite{aizu,arlt,salje} are crystalline solids that undergo a 
shape-changing, structural phase transition with decreasing temperature $T$. 
The high-$T$ or {\em parent} phase distorts spontaneously at the transition 
temperature $T_c$ to form one or more {\em products} or {\em variants} with 
identical energies but different orientations. 
In cubic-tetragonal (C-T) ferroelastics for example, only one of three 
four-fold axes is retained below $T_c$, giving three variants. 

Due to constraints imposed by neighbouring material, a single variant is 
found only rarely, perhaps only in very small grains; 
the displacement resulting from a strain of $10^{-4}$ in a grain of size 
$10\,\mu$m (both typical values) is an order of magnitude too large to be 
accommodated by atomic rearrangements at grain boundaries. 
Multiple variants can arise also from independent nucleation events. 
An external stress can move the walls that separate the variants, so 
converting one variant to the other(s) at little cost in energy; 
the process has been observed for example by neutron diffraction of 
zirconia\cite{kisi01}. 
Related phenomena are strongly hysteretic stress-strain relations, 
shape-memory effects, etc\cite{salje}.

Pockets of the parent phase can persist below $T_c$, until the gain in 
condensation energy overcomes the cost of introducing domain walls. 
Compositional inhomogeneities combined with a strong dependence of $T_c$ on 
composition can also smear the transition, in cases to over 100 K; 
it is common to speak instead of a transformation. 

Domain patterns in ferroelastics have nothing in common with those in 
conventional order-parameter systems. 
Two-dimensional (2d) patterns in tetragonal-orthorhombic (T-O) ferroelastics 
like YBa$_2$Cu$_3$O$_{7-\delta}$ resemble not at all those in 2d Ising models, 
though both have two variants; 
neither can 2d patterns in hexagonal-orthorhombic (H-O) ferroelastics be 
understood from 2d three-state Potts models, both with three.
The difference arises because 
(a) in order to maintain a coherent interface (no dislocations or 
disclinations), ferroelastic domain walls rotate the variants as well as 
join them, and 
(b) order-parameter strains alone are insufficient to understand the patterns.
Disclinations, which have no counterpart in Ising and Potts systems, are 
generated however by wall collisions; 
they dominate patterns especially in H-O-like systems (e.g. 
trigonal-monoclinic lead orthovanadate).  

Much of ferroelastic theory follows Barsch and Krumhansl\cite{barsch84} in 
expanding the free-energy density in powers of the strains and their 
derivatives. 
%CHANGE START 
%CHANGE END
Analytical results are possible in a few cases\cite{barsch84,jacobs85}, but 
structures formed by colliding walls require numerical effort for their 
understanding. 
Such studies\cite{curnoe2,jacobs00,curnoe3} in 2d gave static and transient 
domain patterns that reproduced nearly all aspects of those observed in 
H-O-like and T-O ferroelastics. 
Ref.\onlinecite{lookman03} obtained similar results. 

An alternative approach, phase-field theory\cite{khach}, predates 
Ref.\onlinecite{barsch84} and has also been used extensively to understand 
domain patterns in H-O-like materials\cite{wen}, 
C-T materials\cite{khach2,khach3}, etc. 
Ref.\onlinecite{kartha} discusses differences between the two approaches. 
And other lines have been pursued. 

%CHANGE START 
In C-T ferroelastics, domain walls joining two tetragonal variants lie 
optimally in the cubic 110 planes\cite{sapriel} and are then twin walls. 
In the solutions\cite{barsch84,barsch93,curnoe} for the twin wall, the 
strains depart only locally from their bulk values, decaying exponentially 
with distance from the wall. 
%CHANGE END 
Of course domain walls need not lie in the cubic 110 planes; 
they are then not twin walls, and the departures decay 
algebraically\cite{pertsev00}. 

Optical and electron microscopy of 
%CHANGE START 
zirconia\cite{haya1,haya2,baith}, leucite\cite{palmer,salje}, 
barium titanate\cite{arlt} and other\cite{otherct} C-T ferroelastics 
reveals a variety of patterns. 
The structures\cite{arlt} of small and large grains are respectively lamellar 
(with only two variants) and banded (with all three); 
the latter seems unique to C-T materials. 
%CHANGE END 
The lamellar $\leftrightarrow$ banded transition is 
understood\cite{arlt,pertsev92}, though the analysis is apparently limited 
to variants oriented at $\pi/2$. 
Other aspects of the domain patterns have also been explained 
analytically\cite{haya1,haya3,arlt,pertsev92}. 
The character of the patterns depends also on the thickness of the sample 
and on whether the surface examined is part of clamped specimen, or a free 
surface, or representative of the bulk\cite{arlt,pertsev92}. 
The highly sensitive technique of birefringence imaging\cite{kreisel} 
promises to reveal further details of these patterns. 

The following presents results from simulating the time evolution of C-T 
materials. 
We obtain the equations of motion by expanding the free-energy density and the 
Rayleigh dissipation density to lowest possible order in the strains and 
their derivatives. 
We solve numerically for the displacement, using periodic boundary conditions 
and omitting the inertial term. 
Our late-time structures reproduce most features of the banded patterns found 
in C-T materials\cite{arlt,salje,haya1,haya2,baith,palmer,otherct}.
These structures were not found in the smaller systems used in the previous 
C-T simulations of Refs.\onlinecite{luskin} and\onlinecite{rasmus}; 
some were found in a phase-field study\cite{khach3}. 
Our transient structures show in addition other wall configurations observed 
experimentally. 

\section{Landau theory and equations of motion of elastics}

The displacement $\bfu(\bfr)$ of a material point is defined relative to 
its position $\bfr$ in the parent phase. 
The symmetric strain tensor in this Lagrangian description is 
$\eta_{ij} = \frac{1}{2}(u_{i,j}+u_{j,i}+u_{k,i}u_{k,j})$, 
where $u_{i,j} = \partial u_i/\partial x_j$. 
We neglect the nonlinear term in $\eta$ because it has no known qualitative 
effect on the domain patterns. 
The strains (all of which vanish in the parent phase) are defined by 
\begin{eqnarray}
e_1 & = & u_{1,1} + u_{2,2} + u_{3,3} \ ,\\
e_2 & = & {\textstyle{1\over 2}}           (u_{1,1}-u_{2,2})\ , \\
e_3 & = & {\textstyle{1\over{2\sqrt{3}}}}  (u_{1,1}+u_{2,2}-2u_{3,3})\ ,\\
e_4 & = & {\textstyle{1\over 2}}           (u_{2,3}+u_{3,2})\ ,\\
e_5 & = & {\textstyle{1\over 2}}           (u_{3,1}+u_{1,3})\ ,\\
e_6 & = & {\textstyle{1\over 2}}           (u_{1,2}+u_{2,1})\ ;
\end{eqnarray}
these definitions differ slightly from those in Ref.\onlinecite{curnoe}.
The deviatoric strains $e_2$ and $e_3$ form the two-component order parameter 
of the transition. 
The other strains $e_1$ (the dilatational strain in the small-strain limit) 
and the shear strains $e_4$, $e_5$ and $e_6$ are identically zero in the 
uniform product phase; 
they are required however to understand domain patterns, even for a single 
twin wall\cite{barsch84,barsch93,curnoe}. 
 
The six strains are obtained from the three components of the displacement
${\bf u}$ and so are not independent when they vary spatially;
the second derivatives of the strains are linked by {\em compatibility
relations}, necessary and sufficient conditions that the strains be derivable
from $\bfu$. 
We satisfy these relations implicitly by working with the components $u_i$. 
Refs.\onlinecite{shenoy99,rasmus,lookman03}, which work directly with the 
strains, satisfy them explicitly by imposing them to obtain nonlocal relations 
between the order-parameter strains $e_2$ and $e_3$;
the anisotropic, oscillatory nature of the kernels, obtained also in 
Ref.\onlinecite{kartha}, provides much insight into domain structures and their 
relaxation.
Nonlocal relations were developed much earlier\cite{khach}, though there 
appear to be differences\cite{kartha}. 

In the Landau expansion of the free-energy density in the strains and their 
derivatives, the cubic symmetry of the parent phase permits three invariants 
to second order in the strains, $e_1^2$, $e_2^2+e_3^2$ and 
$e_4^2+e_5^2+e_6^2$. 
The corresponding stiffness coefficients $A_1$, $A_2$ and $A_4$ are linear 
combinations of the Voigt coefficients.
The order-parameter stiffness $A_2$ softens with decreasing $T$ as
$A_2=\alpha(T-T_0)$.
To describe the phase transition, one adds a term cubic in $e_2$ and $e_3$ 
(this term breaks the rotational symmetry in $(e_2,e_3)$ space), and a 
quartic term for stability.
The minimal density, that contains only essential terms, is
\begin{eqnarray}
{\cal F} & = & \frac{A_1}{2}e_1^2 + \frac{A_2}{2}\left(e_2^2+e_3^2\right)
-\frac{B_2}{3}\left(e_3^3-3e_2^2e_3\right)\nonumber\\
& &  +\frac{C_2}{4}\left(e_2^2+e_3^2\right)^2
+\frac{A_4}{2}\left(e_4^2+e_5^2+e_6^2\right)\nonumber\\
& &
+\frac{D_2}{2}\left[\left(\nabla e_2\right)^2 +\left(\nabla e_3\right)^2\right]
\ ;\end{eqnarray}
the last term gives the domain-wall energy (which prevents the system from 
dividing into arbitrarily small domains).
We omit all unnecessary terms, namely higher-order terms and also some of the 
same order as those kept (a second invariant in the order-parameter 
derivatives\cite{curnoe,rasmus} and other derivative invariants).

The coefficients $A_2$, $B_2$ and $C_2$ determine the transition temperature
$T_c$ and the spontaneous strain $e_{30}$ in the product phase. 
When $A_2> B_2^2/4C_2$, the free energy has only the cubic minimum at
$e_2=e_3=0$. 
For $A_2< B_2^2/4C_2$, it has also three degenerate tetragonal minima located
at
\begin{eqnarray}
e_2=0                \ ,\ e_3&=& e_{30}   \\
e_2=-\sqrt{3}e_{30}/2\ ,\ e_3&=&-e_{30}/2 \\
e_2= \sqrt{3}e_{30}/2\ ,\ e_3&=&-e_{30}/2
\end{eqnarray}
with (we assume $B_2>0$) 
\begin{equation}
e_{30}=\frac{B_2+\left(B_2^2-4A_2C_2\right)^{1/2}}{2C_2}.
\end{equation}
The phase transition, which is first-order, occurs when 
$A_2 = \frac{2}{9}B_2^2/C_2$. 
The cubic phase is unstable for $A_2<0$. 

The symmetric stress tensor $\sigma_{ij}$ 
is defined by 
\begin{equation}
\sigma_{ij}=\delta\cf/\delta\eta_{ij} = \cg_k\delta e_k/\delta\eta_{ij}
\end{equation}
with $\cg_k\equiv\delta\cf/\delta e_k $;
explicitly, 
\begin{equation}
\label{g2}
\cg_2=\left(A_2-D_2\nabla^2\right)e_2
                 +2B_2e_2e_3 +C_2e_2\left(e_2^2+e_3^2\right)
\end{equation}
\begin{equation}
\label{g3}
\cg_3=\left(A_2-D_2\nabla^2\right)e_3
 +B_2\left(e_2^2-e_3^2\right)+C_2e_3\left(e_2^2+e_3^2\right)
\end{equation}

\begin{eqnarray}
\sigma_{11} & = &A_1e_1 + {\textstyle{1\over 2        }} \cg_2 
                        + {\textstyle{1\over 2\sqrt{3}}} \cg_3 \\
\sigma_{22} & = &A_1e_1 - {\textstyle{1\over 2        }} \cg_2 
                        + {\textstyle{1\over 2\sqrt{3}}} \cg_3 \\
\sigma_{33} & = &A_1e_1 - {\textstyle{1\over  \sqrt{3}}} \cg_3 \\
\sigma_{23} & = &A_4e_4 \\
\sigma_{31} & = &A_4e_5 \\
\sigma_{12} & = &A_4e_6
\end{eqnarray}

Stresses arise also from dissipative mechanisms.
The same symmetry considerations as used for the free energy give the
Rayleigh dissipative density as
\begin{equation}
{\cal R}  =  \frac{A_1'}{2}\dot{e}_1^2 + \frac{A_2'}{2}(\dot{e}_2^2
+\dot{e}_3^2) +\frac{A_4'}{2}(\dot{e}_4^2+ \de_5^2 +\de_6^2)
\end{equation}
to lowest order in the time derivatives $\de_j$ of the strains. 
The dissipative stresses $\sigma'_{ij}$ are found from 
\begin{equation}
\sigma'_{ij} = \delta {\cal R}/\delta \dot{\eta}_{ij}
= \cg_k' \delta \dot{e}_k/\delta\dot{\eta}_{ij}
\end{equation}
where $\cg_k' \equiv \delta {\cal R}/\delta \de_k$: 
$\cg_1'=A_1' \de_1, \cdots ,\cg_6'=A_4' \de_6$. 

Our interest is in static states and in states where walls move slowly (rather 
%CHANGE START 
than in 
%CHANGE END
effects associated with motion at or near the sound velocity) and so 
we assume isothermal conditions. 
The equations of motion follow from Newton's second law 
\begin{equation}
\label{newt} 
f_i = \rho \ddot{u}_i= \sigma'_{ij,j} + \sigma_{ij,j}\ . 
\end{equation}
In the overdamped limit, the inertial term $\rho \ddot{u}_i$ is dropped and 
Eq.(\ref{newt}) simplifies to $\sigma'_{ij,j} =- \sigma_{ij,j}$.
In terms of the displacement, this is 
\begin{eqnarray}
\label{motion}
& & \hspace{-.3in} \left[A'\partial_1^2 +B'\left(\partial_2^2+\partial_3^2\right)\right] \du_1
     +C'\left(\partial_1\partial_2 \du_2 +\partial_1\partial_3 \du_3\right)
\nonumber \\
&  = &  -\left(A-{\textstyle{1\over3}}D_2\nabla^2\right)\partial_1^2 u_1
        -B\left(\partial_2^2+\partial_3^2\right) u_1
\nonumber \\
 & &     -\left(C+{\textstyle{1\over6}}D_2\nabla^2\right)
           \left(\partial_1\partial_2 u_2 +\partial_1\partial_3 u_3\right)
+R_1^{NL}
\end{eqnarray}
for $i=1$, with obvious forms for $i=2,3$. 
The coefficients are $A'=A_1'+{1\over3}A_2'$, $B'={1\over4}A_4'$ and 
$C'=A_1'-{1\over6}A_2'+{1\over4}A_4'$; 
the definitions for $A$, $B$ and $C$ are obtained by dropping the primes. 
The nonlinear terms 
\begin{equation}
R_1^{NL}=-\partial_1
     \left( {\textstyle{1\over2}}\cg_2^{NL}
           +{\textstyle{1\over2\sqrt3}}\cg_3^{NL}\right)
\end{equation}
\begin{equation}
R_2^{NL}=-\partial_2
     \left(-{\textstyle{1\over2}}\cg_2^{NL}
           +{\textstyle{1\over2\sqrt3}}\cg_3^{NL}\right)
\end{equation}
\begin{equation}
R_3^{NL}=-\partial_3 \left(
           -{\textstyle{1\over \sqrt3}}\cg_3^{NL}\right)
\end{equation}
on the right-hand sides involve the nonlinear parts of $\cg_2$ and $\cg_3$, 
namely the terms with coefficients $B_2$ and $C_2$ in Eqs.(\ref{g2}) and 
(\ref{g3}). 

The matrix on the left-hand side of Eq.(\ref{motion}) must be invertible.
Special handling is required when the strains are constant (that is 
$\partial_1=\partial_2=\partial_3 = 0$) since both sides are then zero. 
Examining the full equation of motion (\ref{newt}), we see that the proper way 
to account for the constant strains is to leave them constant at all times. 
The interpretation is physical: a piece of strained material does not move 
unless there is a differential strain.
Another case of interest is $A_1'=A_4'=0$, which may be a reasonable choice 
given that it is the order parameter, and not the other strain components, 
which changes most quickly in time.
Then the matrix is not invertible for
$\partial_1=0$ or  $\partial_2=0$ or $\partial_3=0$,
but again Eq.(\ref{newt}) tells us how to handle this case.

The equations of motion (\ref{motion}) can be solved under a variety of 
boundary conditions. 
Wishing to examine domain patterns, we used periodic boundary conditions 
(which allow no length change in any direction) in order to force domain 
walls into the low-$T$ phase; 
all three variants are required since two variants can satisfy the constraint 
in only two directions. 
A second important consideration is that these conditions allow use of the 
fast Fourier transform, which is much faster than real-space methods. 
We should however voice our concern that these conditions may lead to spurious 
correlations between relaxation events at large relative distances. 
Of other choices, clamped conditions (${\bf u}=0$ on and outside the 
boundaries, as in Ref.\onlinecite{jacobs00}) would also force domain walls 
into the low-$T$ phase, but are less attractive because they usually give 
complex structures near the edges. 
Open conditions are not useful for our purpose, for the system would 
go to a single variant for almost any initial state.
Yet another possibility corresponds to applied stresses at the boundaries. 

%CHANGE START 
In Fourier space, Eqs. (\ref{motion}) are identical to the equations of 
time-dependent Ginzburg-Landau (TDGL) theory\cite{shenoy99,rasmus,lookman03}. 
%CHANGE END 
In real space, Eqs. (\ref{motion}) contain extra space derivatives on both 
sides relative to TDGL theory and so appear more general (they can be applied 
for example to systems with imposed strains). 
The neglect of the inertial terms both here and in 
Refs.\onlinecite{shenoy99,rasmus,lookman03} is problematic for the relaxation. 

\section{Domain patterns}
We solved Eqs.(\ref{motion}) numerically, as described in the Appendix, 
starting usually from random displacements. 
We present fully converged\cite{fulcon} results for 3d grids of $128^3$ 
points and quasi-static\cite{fulcon} results for $256^3$ points. 
We present also transient structures in systems of $256^3$ points. 
Systems of these sizes reveal features not seen in previous studies, which 
used at most $64^3$ points. 

The nature of the patterns depends in part on the stiffness coefficients 
$A_1$ and $A_4$. 
The dilatational and shear energies are minimized when the walls lie in cubic 
110 planes, and so the parameters $A_1$ and $A_4$ control the energy cost 
incurred when the walls depart from their optimal orientations. 
In stiff systems, with large $A_1$ and $A_4$, the domain walls must lie close 
to the 110 planes, whereas in soft systems they can depart from these optimal
orientations at small cost in energy.
We have however no quantitative way to distinguish stiff from soft systems;
comparing $A_1$ and $A_4$ with the order-parameter stiffnesses\cite{curnoe2}
found from the curvatures of the free energy about the tetragonal minima is
not an effective means.

\subsection{Late-time structures}
We made extensive studies of late-time domain structures at $A_2=-2$ and 
$A_2=-20$; 
the spontaneous strain $e_{30}$ at these temperatures is respectively twice 
and four times the value at $T_c$. 
We present results at only the lower (latter) value, where the order 
parameters inside the domains are more well developed. 

Most of $\approx200$ simulations gave multiply banded or herringbone 
structures like those shown in Figure 1\cite{startfig1}. 
These patterns, which seem at first glance to reflect more the periodic 
boundary conditions than any physics, are in fact found in the microstructure 
of polydomain zirconia (examples are Figures 4(a), 5(a), 6 and 7 of 
Ref.\onlinecite{haya1}, Figure 3 of Ref.\onlinecite{haya2}, and Figures 1 and 
2 of Ref.\onlinecite{baith}) and other C-T materials\cite{otherct}. 
Similar patterns appear also in the well known ferroelectric BaTiO$_3$, 
specifically Figures 2(a) and 8(b) of Ref.\onlinecite{arlt}; 
they should appear in other elastic/electric and elastic/magnetic ferroics 
provided that the elastic energy dominates the electric and magnetic energies, 
as it does\cite{arlt} in BaTiO$_3$.  
Banded structures were found also in a phase-field study\cite{khach3}. 

All three variants appear in Figure 1, but not equivalently. 
The structures consist of two primary bands, here red and green; 
the width of the primaries is determined by the system size in Figure 1 
and, one assumes, by the grain size in experiment. 
Each primary is penetrated by the third variant, here blue; 
neither primary contains domains of the other. 
Within each primary, the host and the third variant form secondary bands; 
the ratio of the width of the host variant to the width of the third variant 
in the secondary bands is ideally 2:1 so that the three variants appear with 
equal volume fractions\cite{haya1,haya3}. 
The same ratio was found in Refs.\onlinecite{arlt,pertsev92}, which found 
also the optimal value of the period of the primary bands to that of the 
secondary bands. 

%number 215f
Figure 1(a) shows a fully converged\cite{fulcon} structure; 
it has the lowest energy of seven states found in 24 quenches with the same 
parameters. 
The front face shows chevron (or herringbone) structures; 
the blue domains are continuous across the red-green boundaries, where they 
bend through $90^\circ$ and are slightly distorted as well. 

%number 211g
Figure 1(b) shows another fully converged structure obtained with the same 
parameters as part (a); 
it has a higher energy (the third lowest of the seven states) and so is 
metastable. 
Some of the blue variants form chevrons, as in part (a), but some terminate
at the red-green boundaries;
the walls occasionally deviate from the 110 planes.
Presumably the higher energy relative to part (a) results in part from the 
terminations and the deviations; 
both are seen experimentally, for example in Figures 4(a), 5(a) and 6 of 
Ref.\onlinecite{haya1}.

%number 250p
Figure 1(c), for a $256^3$ system with stiffer parameters, shows a 
quasi-static\cite{fulcon} configuration with a mixture of continuing and 
terminating variants.
The system is large enough to show the secondary banding clearly, but it is 
too small and has too many imperfections to display well the 2:1 ratio 
discussed above. 
Of the two other simulations performed with the same parameters, one gave no 
terminations and not surprisingly a lower energy, and the third gave more 
terminations and a larger energy. 

In addition to these banded structures, we found lamellar structures in 
smaller systems ($64^3$), particularly for stiffer parameters; 
the agreement with the lamellar $\leftrightarrow$ banded transition analysed in 
Ref.\onlinecite{arlt,pertsev92} is however largely illusory, for the order 
parameters cannot approach their optimal values due to the length constraint 
in the third direction. 
We found also intermediate structures in which the narrow variant appears in 
only one of the two primary bands. 
Finally, a few systems gave very different tweed-like or basket-weave 
structures, of all three variants, that seem not to be observed in experiment. 

\subsection{Transient structures}
In all our late-time banded structures (not just those of Figure 1), walls 
collide only at boundaries between the primary bands. 
Within each primary red (green) band, 
\vskip 0.001 true cm
\noindent
(a) the red-blue (green-blue) walls adopt only one of the two possible 
orthogonal orientations\cite{sapriel}, and 
\vskip 0.001 true cm
\noindent
(b) the other primary, the green (red) variant, is absent. 
\vskip 0.001 true cm

Walls colliding within the primary bands are however observed; 
examples are the A$_3$ band in Figure 6 of Ref.\onlinecite{haya1}, 
Figure 7(b) of Ref.\onlinecite{haya1}, 
Figure 3 of Ref.\onlinecite{haya2}, and 
Figure 11 of Ref.\onlinecite{baith}. 
Because we obtain only two primary bands, neither did we observe the A$_2$ and 
A$_4$ bands in Figure 6 of Ref.\onlinecite{haya1}; 
these contain the same two variants as the A$_1$ band but with the orthogonal 
wall orientation. 

Perhaps our systems are too small to show these effects, perhaps the 
experimental systems are incompletely relaxed; 
that different experimental conditions can give different patterns\cite{arlt} 
may be relevant here. 
We have however found some of these features in transient structures of a 
$256^3$ system; $128^3$ systems are too small to show interesting features 
clearly. 

The important aspects of Figures 2(a) and (b) are the following. 
\vskip 0.001 true cm
\noindent
1. Needle twins: 
The top face of part (a) shows a band of green needle twins in the red primary 
band, and also a band of blue needle twins in the green primary band; 
needles appear also elsewhere. 
In part (b), some needles have advanced and some have retracted. 
Needle twins are found for example in leucite (Figure 3 of 
Ref.\onlinecite{palmer}). 
\vskip 0.001 true cm
\noindent
2. Collisions of identical variants: 
In the green primary band on the top face, green/blue walls collide with 
green/blue walls of the other orientation, forming modulated structures. 
Figure 11 of Ref.\onlinecite{baith} shows a similar pattern in zirconia, 
``a rather exceptional case'', also formed by orthogonal colliding walls. 
As in Refs.\onlinecite{jacobs00,curnoe3}, we ascribe these modulations, and 
also the structures in Figure 3(b) of Ref.\onlinecite{palmer}, to formation of 
wedge disclinations between two identical but differently rotated variants; 
the same explanation applies to tip splitting, in some cases. 
\vskip 0.001 true cm
\noindent
3. A split tip: 
Tip splitting seems to occur only rarely in C-T materials (relative to T-O 
materials), presumably because of the extra freedom afforded by three variants; 
an example is Figure 3 of Ref.\onlinecite{palmer}. 
We found only one split tip, an indistinct one at that, at the right side 
of the top face in part (a); 
this is of course a transient configuration. 
Split tips appear in the statics of some simulations\cite{jacobs00,khach3}, 
but only at the interface with parent material; 
they are more frequent in transient structures\cite{curnoe3}. 
\vskip 0.001 true cm
\noindent
4. Collisions of different variants: 
In the top faces of both parts (a) and (b), green needle twins collide 
orthogonally with, or come near, blue variants; 
collisions occur also in the lower front face of part (a). 
No special features result from the collisions, which have not been noted in 
any experiment known to us. 

Figure 2(c) shows the same system at the later time $t=30$. 
Many of the defects in part (b) have disappeared in this quasi-static wall 
configuration; 
the walls are straighter, but in this relatively soft system they still bend 
where the third variant (here blue) terminates. 

Inspection of the dilatational and shear strains of late-time structures 
shows, not surprisingly, that their magnitude is maximum in the wall-collision 
regions. 
We investigated also the early stages of growth initiated by locally 
perturbing the parent phase at a temperature well below $T_c$; 
the growth occurs predominantly along spikes in the 111 directions and 
planes in 110 directions, producing a noncompact object. 

\acknowledgments
Some of the numerical simulations were performed at the Computation and 
Visualisation Centre at Memorial University of Newfoundland. 
We are grateful to E. H. Kisi for pointing out the similarity of our patterns 
to those observed in zirconia and for informing us of 
Refs.\onlinecite{haya1,haya2,baith} and\onlinecite{haya3}, 
to J. Kreisel for informing us of Ref.\onlinecite{kreisel}, 
to T. Hahn for informing us of Ref.\onlinecite{arlt}, 
to M. Hayakawa for discussions, 
and to NSERC of Canada for financial support.

\appendix
\section{}

We solved Eqs.(\ref{motion}) using periodic boundary conditions on the 
displacement ${\bf u}$.
Since only qualitative comparison with experiment seems possible at present, 
and because we wished to obtain results qualitatively applicable to many 
C-T materials, we scaled the energy, the strains and the length; 
this scaling requires neglect of the nonlinear term in the strain tensor
$\eta$. 
We chose the values $B_2=3\times10^3$ and $C_2=2\times10^6$ so that the 
transition occurs at $A_2=1$, and the scaled strain at $T_c$ is 
$e_{30}(T_c)=10^{-3}$, an arbitrary value.
We chose $D_2= 1$ to set the scale for the domain-wall width\cite{curnoe}. 
The number of parameters in the free-energy density is then reduced from 
six to three, the scaled stiffnesses $A_1$ and $A_4$ and the scaled $T$-like 
variable $A_2$. 
Results for other parameter values are easily obtained by scaling back to the 
original variables. 

The viscosity coefficients $A_j'$ appearing in Eq. (\ref{motion}) are not 
known from microscopic theory. 
Neither we expect can they be determined from experiments such as ultrasonic 
attenuation that operate on time scales very different from those governing 
domain-wall motion (as distinct from the ``twin cry'' in some materials).
Lacking experiments that might determine the relevant coefficients, we chose 
the unit of time so that $A_2'=1$;
lacking a reason to do otherwise, we chose $A_1'=A_4'=1$ also.

Each time step began with the Fourier coefficents
$\tilde u_j({\bf k},t)$ at time $t$.
The left-hand sides of Eqs. (\ref{motion}) are linear in the strains and so 
the space derivatives are obtained by multiplication in Fourier space;
our approximations for the derivatives are described below.
The linear terms on the right-hand sides are found in the same way.
To obtain the nonlinear terms $R_i^{NL}$ in Fourier space, we formed the
Fourier coefficients of the strains $e_2$ and $e_3$, transformed them to find
the strains in real space, found the nonlinear terms ${\cal G}_2^{NL}$ and
${\cal G}_3^{NL}$ by multiplication (replacing the strains point by point),
transformed the two terms back to Fourier space, and multiplied to obtain the
space derivatives in Fourier space.
The solutions were advanced in time by an Euler step (usually 
$\Delta t = 4\times10^{-3}$); 
solution of three linear algebraic equations then gives the three 
components $\tilde u_j({\bf k},t+\Delta t)$ in Fourier space.
To monitor the convergence, we found the energy, the root-mean-square
order-parameter strains and right-hand sides of Eq.(\ref{motion}), etc,
every 10 or 20 time steps.

The above computational scheme requires storage of five matrices, three for 
the $\tilde u_j({\bf k},t)$ and two for the strains $e_2$ and $e_3$ (or 
${\cal G}_2^{NL}$ and ${\cal G}_3^{NL}$).
The fast Fourier transforms were performed using the Numerical Recipes
routine {\tt fourn}\cite{numrec}, which deals with complex matrices.
The full executable file for a $256^3$ system requires 1.35GB of storage.
Savings of about two in storage and execution time would be obtained by use of
routines for real matrices, at though some expense in coding and clarity.
We have also obtained static structures by conjugate-gradient minimization 
of the energy\cite{numrec}. 
%CHANGE START 
The latter method is preferable in some respects to solving the equations of 
motion, and we have used it to verify the correctness of some of our 
$64^3$ and $128^3$ results; 
it requires however roughly 7 times more storage, well in excess of that 
%CHANGE END
available to us for $256^3$ systems. 

The first and second derivatives were obtained from obvious generalizations
of the 1d finite-difference approximations
\begin{equation}
{df(0)\over dx} \approx {2\over 3h} \left\{
 \left[f\left(h\right)- f\left(-h\right)\right]
   -{1\over8}
 \left[f\left(2h\right)- f\left(-2h\right)\right]
\right\}
\end{equation}
\begin{eqnarray}
{d^2 f(0)\over dx^2} &\approx & {4\over 3h^2} \left\{
 \left[f\left(h\right)+ f\left(-h\right)
                      -2f\left(0\right) \right]  \right.\nonumber \\
& &  \left.   -{1\over16}
 \left[f\left(2h\right)+ f\left(-2h\right)
                       -2f\left(0\right) \right]
\right\}
\end{eqnarray}

If the Fourier coefficients are defined by
\begin{equation}
f(\bfr)=\sum_{jkl}a_{jkl} e^{2\pi i(jx+ky+lz)/L}
\end{equation}
with period $L=Nh$ in each variable, then the first and second derivatives
are approximated by
\begin{equation}
{\partial \over \partial x} \to {i\over 3h} \sin(2\pi jh/L)
               \left[3+2\sin^2\left(\pi jh/L\right)\right]
\end{equation}
\begin{equation}
{\partial^2 \over \partial x^2} \to {-4\over 3h^2} \sin^2(\pi jh/L)
               \left[3+ \sin^2\left(\pi jh/L\right)\right]
\end{equation}
in Fourier space.
In obtaining second-derivative terms like $\partial_1^2$ in
Eq.(\ref{motion}), it is important to use Eq.(A5) rather than Eq.(A4) twice,
for the latter vanishes at $j=N/2$.

The space step size $h$ must be chosen as a reasonable compromise between
the conflicting demands of large physical size $L=Nh$ on the one hand and
accuracy on the other.
Our values  $h=0.5$ at $A_2=-2$ and $h=0.25$ at $A_2=-20$ were established 
as follows. 

We first performed 48 quenches at $A_2=-20$ with $A_1=A_4=100$ on systems of
identical linear size $L=Nh$;
24 of these quenches used $(N,h)=(128,0.125)$ and 24 used $(N,h)=(64,0.25)$.
The larger step size gave 4 states, each of which was clearly identifed with
a state found for the smaller $h$.
The energies of the 4 states common to the two sets of quenches agreed to 
better than 1 part in 2000 (relative to the uniform product phase) and the 
relative frequencies of occurrence were comparable.
The smaller step size gave however 3 additional states, each once.
Since $h=0.25$ is satisfactory at $A_2=-20$ and since the variational wall 
width\cite{curnoe} scales as $1/e_{30}$, one expects $h=0.5$ to be 
satisfactory at $A_2=-2$ where $e_{30}$ is half the value at $A_2=-20$.
Less extensive tests carried out at $A_2=-2$ with $A_1=A_4=100$ gave
comparable results for $(N,h)=(128,0.25)$ as against $(64,0.5)$.
Similar tests with $A_1=A_4=1000$ at both values of $A_2$ gave the same
conclusions.

In passing, we remark that the number of metastable states found in the 
quenches of the previous paragraph and in the quenches used for parts (a) 
and (b) of Figure 1 is smaller than found in a T-O study\cite{jacobs00} of 
clamped systems of comparable linear size; 
the difference is due to the different boundary conditions. 

\begin{figure}
\caption{
%Fig. 1 caption:
%215f, 211g, 250p
Late-time domain patterns. 
The faces of each display cube are cubic 100 planes; 
the red, green and blue regions correspond to the three tetragonal variants;
the domain walls (black) lie optimally in cubic 110 planes. 
Parts (a) and (b) show fully converged\cite{fulcon} patterns (both at 
$t\approx70$) obtained in $128^3$ systems using different starting 
configurations but otherwise identical parameters ($A_1=A_4=100$); 
pattern (b) is a metastable state with higher energy than pattern (a). 
Part (c) shows a quasi-static\cite{fulcon} configuration in a $256^3$ system 
at $t=140$ with stiffer parameters ($A_1=A_4=500$).
The temperature parameter is $A_2 = -20$ for all three parts. 
}
\end{figure}

\begin{figure}
\caption{
%Fig. 2 caption: 
% 26222, 26236 and 262f 
Snapshots of a $256^3$ system as it relaxes from random initial displacements. 
The faces are cubic 100 planes. 
The temperature parameter is $A_2=-20$ and the stiffnesses are $A_1=A_4=100$. 
Parts (a) and (b) are transient structures at times $t=2.2$ and 3.6 
respectively; 
part (c) is a quasi-static pattern at $t=30$. 
}
\end{figure}


\begin{references}
\bibitem{fulcon} 
Fully converged means that numerical errors prevented further change in the 
energy, root-mean-square strains, etc. 
Quasi-static means that gross features of the pattern are fixed; 
further relaxation occurs, without wall annihilation or creation, through 
wall migration so as to align where possible identical variants across 
boundaries. 
The relaxation times for this process are very large and increase rapidly 
with the system size and.
\bibitem{startfig1}
Whether the narrow domains in the banded structures were red, green or blue 
depended on the initial conditions. 
In preparing the Figures, we used where necessary symmetry operations to make 
the narrow domains blue.
\end{references}
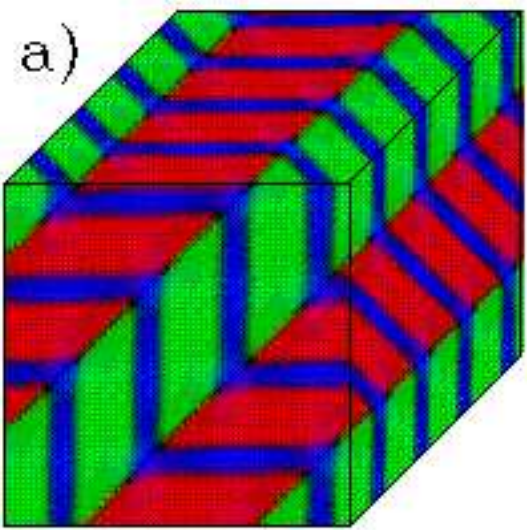
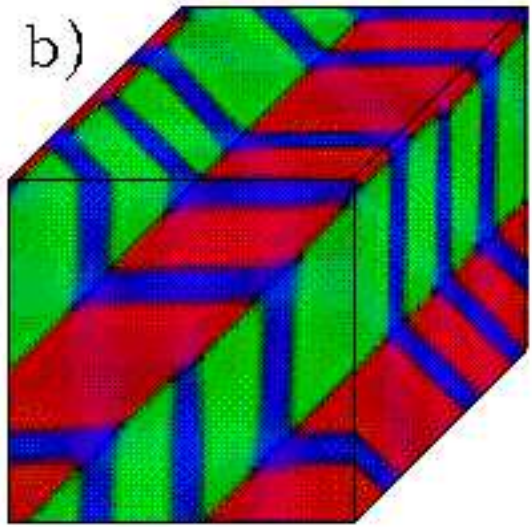
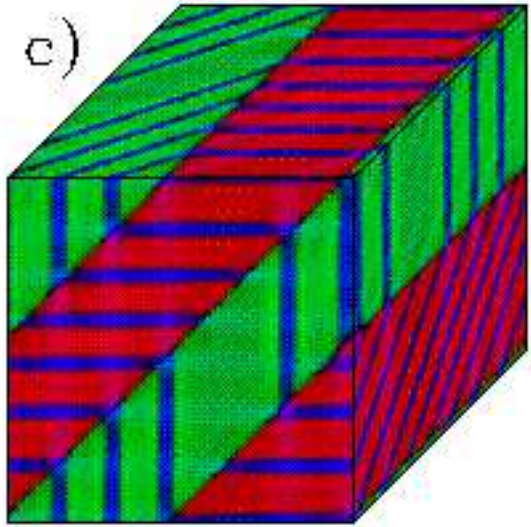
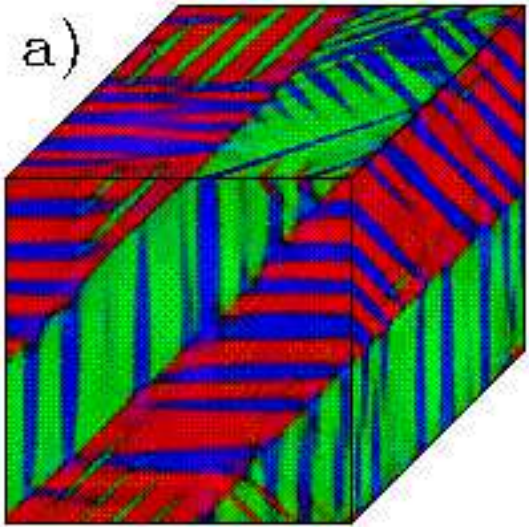
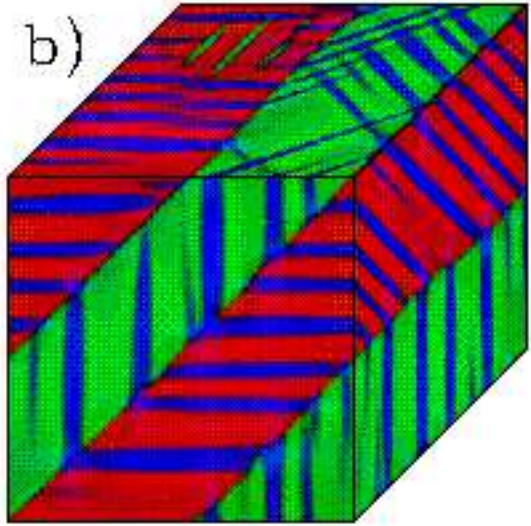
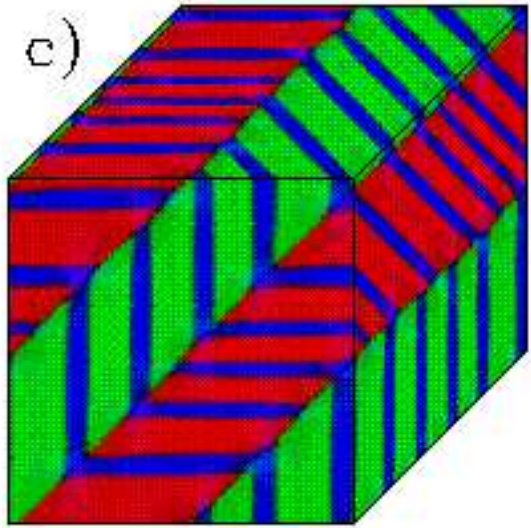
\end{document}